\documentclass[onecolumn,prb,showpacs,manuscript]{revtex4}
\usepackage{dcolumn}
\usepackage{amsmath}
\usepackage{amssymb}
\usepackage[dvips]{epsfig}  % to include PostScript figures
\makeatletter
\def\btt#1{\texttt{\@backslashchar#1}}%
\DeclareRobustCommand\bblash{\btt{\@backslashchar}}%
\makeatother

\begin{document}
%\draft
\title{\bf Potential ultra-incompressible material ReN: first-principles prediction }
\author{Yanling Li$^{1,2,3}$, Zhi Zeng$^{1}$\footnote{Corresponding author. Email address: zzeng@theory.issp.ac.cn}, Haiqing Lin$^{3}$}
\affiliation {1.Key Laboratory of Materials Physics, Institute of
Solid State Physics, Chinese Academy of Sciences, Hefei
230031,People's Republic of China and \\ Graduate School of the
Chinese Academy of Sciences, Beijing 100049, People's Republic of
China\\
2.Department of physics, Xuzhou Normal University, Xuzhou 221116,
People's Republic of China\\
3.Department of Physics and Institute of Theoretical Physics, The
Chinese University of Hong Kong, Shatin, Hong Kong, China.}

\date{April 24, 2008}% It is always \today, today, but you may specify any date with \date.

\begin{abstract}
The structural, elastic and electronic properties of ReN are
investigated by first-principles calculations based on density
functional theory. Two competing structures, i.e. CsCl-like and
NiAs-like structures, are found and the most stable structure,
NiAs-like, has a hexagonal symmetry which belongs to space group
$P6_{3}/mmc$ with \emph{a}=2.7472 and \emph{c}=5.8180 \AA. ReN
with hexagonal symmetry is a metal ultra-incompressible solid and
has less elastic anisotropy. The ultra-incompressibility of ReN is
attributed to its high valence electron density and strong
covalence bondings. Calculations of density of states and charge
density distribution, together with Mulliken atomic population
analysis, show that the bondings of ReN should be a mixture of
metallic, covalent, and ionic bondings. Our results indicate that
ReN can be used as a potential ultra-incompressible conductor. In
particular, we obtain a superconducting transition temperature
$T_c$$\approx$4.8 K for ReN.

\end{abstract}


\maketitle
\section{Introduction}
Transition metal nitrides are of intense interest for
researchers.\cite{Horvath-Bordon,PMcMillan,Zerr,Crowhurst,Young,Soignard,Bull}
Not only do they offer the fundamental scientific challenge of
finding methods to control, tune and enhance their peculiar
physical properties, but they also have important technical
applications such as cutting tools,\cite{Horvath-Bordon,Soignard}
oxidation-resisting materials,\cite{Zerr} optical and magnetic
apparatus.\cite{Crowhurst} Transition metal nitrides possess
particular mechanical, optical, electric, and magnetic properties,
which attribute to their unusual electronic bonding (a mixture of
covalent, metallic, and ionic bondings) and strong electron-phonon
interaction. The higher valence electron density (VED) and the
stronger hybridization between transition metal \emph{d} electrons
and nonmetal \emph{p} electrons as well as less elastic anisotropy
make some of them exhibit large incompressibility and high
hardness, considered as powerful candidates of superhard
materials.\cite{Kaner}

At present, there are three kinds of transition metal nitrides
confirmed: X$_{3}$N$_{4}$ (X=Hf, Zr),\cite{Zerr} YN$_{2}$ (Y=Pt,
Ir, Os),\cite{Crowhurst,Young} and ZN (Z=Ti, Ta, Nb,
Mo).\cite{Soignard,Bull} Many works, including determination of
the structure, discussion of elastic and electrical properties,
have been carried out theoretically and
experimentally.\cite{Crowhurst,Young,Soignard,Bull,Mattesini,Young-1,Chen,Yu,Wu,Montoya,Fan,Kanoun}
The interaction of experiment and first-principles calculation has
become a powerful tool to predict, test and confirm the structure
of the new materials.\cite{Horvath-Bordon} For example, jointing
experiment and first-principles calculation,
OsN$_{2}$,\cite{Young,Chen,Yu,Wu,Fan}
IrN$_{2}$\cite{Crowhurst,Young,Yu} and
$\delta$-MoN\cite{Soignard,Bull,Kanoun} were identified as
orthorhombic $Pnnm$, monoclinic $P2_{1}c$ and hexagonal $P6_{3}mc$
structures, respectively. The sites of nitrogen atoms in unit cell
can be confirmed by theoretical calculation, which makes it
possible to deeply study the mechanical, electronic and magnetic
properties of transition metal nitrides.
%So far few studies on ReN were reported experimentally.
Haq and Meyer\cite{Haq} reported the superconducting and
electrical properties of Re$_{1-x}$N$_{x}$ ($x$=0.13$\sim$0.5) by
means of ion implantation experimental technique. The rhenium
nitride with $fcc$ structure was observed and the corresponding
superconducting temperature $T_{c}$ value of 4.5-5.0 K was
obtained. However, Isaev \emph{et al}\cite{Isaev} pointed out that
transitional metal mononitrides with ten and more valence
electrons in the NaCl-type structure are dynamically unstable
which has been verified in MoN and NbN. \cite{Kanoun,Isaev} As
such, ReN with face centered cubic structure maybe unstable
mechanically. So the crystal structure of ReN remains an open
question to date. Moreover, to the best of our knowledge, the
mechanical property of ReN has not been reported.

In this paper, the structural, mechanical and electronic
properties of ReN are investigated by first-principles
calculations. Because most of the transition-metal mono-nitrides
are known to be generally based on cubic or hexagonal metal
sublattice, and N atoms occupy octahedral or trigonal prismatic
interstitial sites, therefore, all the possible cubic and
hexagonal structures are chosen as candidate structures of ReN,
including zinc blende (ZB) (space group $F\bar{4}3m$), rocksalt
(NaCl) (space group $Fm\bar{3}m$), CsCl (space group
$Pm\bar{3}m$), WC (space group $P\bar{6}m2$), and NiAs types
(space group $P6_{3}/mmc$).\cite{MoN} Then incompressibility and
elastic anisotropy are discussed. Further, the superconduction
transition temperature $T_c$ of ReN is estimated using the
modified McMillan equation by Allen and Dynes.\cite{Allen}
Finally, the electronic property of ReN is given by electron
structure calculation and Mulliken atomic population analysis.

\section{Computational details}
Our calculations on structural and electronic properties of ReN
are based on the full-potential linearized augmented plane waves
plus local orbitals method (FP-LAPW)\cite{APW} as implemented in
the Wien2k package to solve the scalar-relativistic Kohn-Sham
equations.\cite {wien2k} For the exchange correlation energy
functional, the local density approximation (LDA) is employed. The
muffin-tin radii are set to 1.9 and 1.60 bohr for Re and N,
respectively. The maximum value $l_{max}$ for the wave function
expansion inside the atomic spheres is limited to 10. We expand
the basis function up to \emph{R}$_{MT}$\emph{K}$_{max}$=7.0
(\emph{K}$_{max}$ is the maximum modulus of the reciprocal lattice
vector). Full relativistic approximation is used for the core
electrons, and scalar relativistic approximation is used for the
valence electrons of 5\emph{s}5\emph{p}4\emph{f}5\emph{d}6\emph{s}
for Re and of 2\emph{s}2\emph{p} for N. The self-consistent cycle
is achieved by taking 1200 points in the first Brillouin zone. The
convergence has been followed with respect to the energy and
density.

The elastic constants of ReN are obtained within the framework of
the finite strain technique by CASTEP code using first-principles
plane-wave basis pseudopotential method (PW-PP) based on
DFT.\cite{Segall} All the possible structures are also optimized
by the BFGS algorithm\cite{BFGS} which provides a fast way of
finding the lowest energy structure and supports cell optimization
in the CASTEP code. In the calculation, the interaction between
the ions and the electrons is described by using Vanderbilt's
supersoft pseudopotential with the cutoff energy of 310 eV. In the
geometrical optimization, all forces on atoms are converged to
less than 0.002 eV/\AA, all the stress components are less than
0.02 GPa, and the tolerance in self-consistent field (SCF)
calculation is 5.0$\times$10$^{-7}$ eV/atom. Relaxation of the
internal degrees of freedom is allowed at each unit cell
compression or expansion. From the full elastic constant tensor we
determine the bulk modulus $B$ and the shear modulus $G$ according
to the Voigt-Reuss-Hill (VRH) approximation.\cite{Hill} The
Young's modulus, $E$, and Possion's ratio, $\nu$, can be
calculated by the formulae
%\begin{equation}\label{E-nu}\nonumber
$E=\frac{9BG}{3B+G} , \nu= \frac{3B-2G}{2(3B+G)}$.
%\end{equation}

\section{Results and discussions}
%\subsection{Structural property}
The equilibrium volume, bulk modulus and its pressure derivative
are obtained by fitting the total energy calculated at different
volumes to the 3-rd order Birch-Murnaghan equation of
state.\cite{eos} The equilibrium structural parameters, density,
valence electron density, the total energy, as well as the bulk
modulus and its pressure derivative obtained from FP-LAPW method
are listed in Table I. The energy-volume curve is plotted in Fig.
1. ReN is more energetically stable in the hexagonal structure
under ambient pressure than cubic structure and NiAs-ReN structure
is most stable, as shown in Table I and Fig. 1. In NiAs-ReN (space
group $P6_{3}/mmc$) the unit cell contains two chemical formula
units (f.u.) in which two Re atoms are at $2a$ (0, 0, 0) and two N
atoms at $2c$ ($\frac{1}{3}$, $\frac{2}{3}$, $\frac{1}{4}$) sites.
The equilibrium volumes per formula unit (f.u.) are 23.5037
\AA$^{3}$ for ZB-ReN, 19.5258 \AA$^{3}$ for NaCl-ReN, 18.9065
\AA$^{3}$ for CsCl-ReN, 19.4775 \AA$^{3}$ for WC-ReN, and 19.0162
\AA$^{3}$ for NiAs-ReN, which are higher than that of Re metal
(14.7173 \AA$^{3}$)\cite{Neumann,Chung} due to the adding of N
atoms. We note that the expansion of lattice when incorporating
the N in the pure Re is not high enough to compensate for the
decrease for the interatomic distance between the first neighbors.
The nearest-neighbor distance $d$ between atoms in ReN is much
less than that of Re (which is about 2.74 \AA), which means that
ReN maybe possess the much higher bulk modulus than pure Re metal
(discussed later). The densities of ReN are 14.1476, 17.0299,
17.5877, 17.0721, and 17.4862 g/cm$^{3}$, for ZB-ReN, NaCl-ReN,
CsCl-ReN, WC-ReN, and NiAs-ReN, respectively.

The valence electron shell of Re is $5d^{5}$$6s^{2}$, and that of
N is $2s^{2}p^{3}$ so the total valence electron number is 12 per
ReN molecule. The calculated VEDs are 0.5106 electrons/\AA$^{3}$
for ZB-ReN, 0.6146 electrons/\AA$^{3}$ for NaCl-ReN, 0.6347
electrons/\AA$^{3}$ for CsCl-ReN, 0.6162 electrons/$\AA^{3}$ for
WC structure, and 0.6310 electrons/\AA$^{3}$ for NiAs-ReN. It is
worthy to note that all VEDs for these three structures are higher
than that of Re metal (0.4761 electrons/\AA$^{3}$)\cite{Chung} and
are compared with 0.70 electrons/\AA$^{3}$ for
diamond.\cite{Gilman} Generally, the higher VED, the stronger the
incompressibility. In addition, from Table I, it is clear that the
bulk modulus of ReN is larger than that (360 GPa)\cite{Chung} of
Re metal except for ZB-ReN (unstable mechanically). Thus, we
conclude that the ReN is less compressible than the pure Re.
%The almost same in bulk modulus $B_{0}$ computed (Table I) for
%NiAs-ReN and MoN-ReN is easily understood, considering that the
%two structures are closely related.
It is delighted that the estimated bulk modulus of ReN with NiAs,
CsCl, and WC structures is even comparable to that of
diamond,\cite{diamond} which means that the ReN with these
structures has a stronger incompressibility, indicating that ReN
can be a candidate of superhard material.

%\subsection{elastic property}
In order to get more accurate elastic constants of ReN using PW-PP
scheme, all considered structures are optimized
firstly.\cite{optimize} The calculated elastic constants are
presented in Table II. For a stable hexagonal structure, the five
independent elastic constants $c_{ij}$
($c_{11}$,$c_{33}$,$c_{44}$,$c_{12}$ and $c_{13}$) should satisfy
the well known Born-Huang criteria for stability:\cite{Born}
\begin{equation}\label{stability}\nonumber
c_{12}>0, c_{33}>0, c_{11}>c_{12}, c_{44}>0,
(c_{11}+c_{12})c_{33}>2c^{2}_{13};
\end{equation}
while for cubic crystal, the three independent elastic constants
$c_{ij}$ ($c_{11}$,$c_{12}$ and $c_{44}$) satisfy
inequalities:\cite{Born}
\begin{equation}\label{stability}\nonumber
c_{44}>0, c_{11}>\left|c_{12}\right|, c_{11}+2c_{12}>0.
\end{equation}
Clearly, our calculated elastic constants $c_{ij}$ for both cubic
CsCl-ReN and the hexagonal NiAs-ReN satisfy the Born-Huang
stability criteria, suggesting that they are mechanically stable.
Cubic ZB-ReN and NaCl types are unstable mechanically because of
their negative $c_{44}$. Our results stand by for Isaev's
viewpoint of which the transition metal mononitrides with ten and
more valence electrons in the NaCl-type structure are unstable
mechanically. WC-ReN type is also unstable mechanically due to the
very small $c_{44}$ value (near zero). In the following, only CsCl
and NiAs types of ReN are discussed.

From the elastic constants calculated above, bulk modulus $B$,
shear modulus $G$, Young's modulus $E$, and Possion's ratio $\nu$
are obtained and listed in Table II, which are important in order
to understand the elastic properties of ReN. The bulk modulus of
ReN calculated by using elastic constants agrees well with the one
obtained through the fit to the 3rd order Birch-Murnaghan EOS,
which shows that the discussion-above on compressibility of ReN is
reasonable. To further discuss the incompressibility of ReN, the
evolution of the volume compressions as a function of pressure are
plotted in Fig. 2. For comparison, the volume compression of
diamond is given simultaneously. It is evident that ReN with NiAs
structure is slight higher incompressible than diamond, while ReN
with CsCl structure has slightly less incompressible than diamond
in the entire pressure range (Fig. 2(b)).

To explain the directional dependence of compression, fraction
axis compression as a function of pressure for diamond and two
structures of ReN are given in Fig. 3. Compression of the axes
presents interesting anisotropy. For hexagonal NiAl-ReN, the $c$
axis is more incompressible than the $a$ axis. The
incompressibility along $a$-axis is slightly less than that of
diamond, while $c$ axis is even more incompressible than the
analogous axis of diamond. While for cubic CsCl-ReN, it is
slightly lower incompressible than diamond (Fig. 2 and Fig. 3).
Therefore, we can claim that ReN is an ultra-incompressible solid.
The shear modulli obtained are 248 GPa for CsCl-ReN and 237 GPa
for NiAs-ReN, and corresponding $G/B$ values are 0.588 and 0.526.
The Possion's ratio is 0.25 and 0.28 for CsCl and NiAs types,
respectively, which indicates that ReN has central interatomic
forces and is relatively stable against shear.\cite{Ravindran} The
high $G/B$ ratio or, equivalently, the low Poisson's ratio also
implies a high value degree of covalency.

It is well known that the elastic anisotropy of crystals has
significant application in engineering science since it is highly
correlated with both the hardness of materials and the possibility
to induce microcracks in the materials. Hence it is important to
calculate elastic anisotropy in order to understand the properties
and improve the durability of materials. The anisotropy factor for
cubic CsCl-ReN, $A$=$(2c_{44}+c_{12})/c_{11}$=0.45, shows that
cubic CsCl-ReN is elastic anisotropy. The anisotropy of linear
compression along the $a$ axis and $c$ axis with respect to the
$b$ axis can be evaluated by means of two parameters: $A_{B_{a}}$
and $A_{B_{c}}$,\cite{Ravindran} where a value of 1 indicates
compression isotropy and any departure from one corresponds to a
degree of compression anisotropy. For low symmetry crystal, the
percentage of anisotropy in compressibility and shear can be
derived from two expressions, bulk modulus anisotropic factor
$A_{B}$ and shear anisotropy factor $A_{G}$,
respectively.\cite{Ravindran} The calculated linear compression
parameters $A_{B_{a}}$, $A_{B_{c}}$ and two other anisotropy
factors $A_{B}$ and $A_{G}$ are listed in Table III. NiAs-ReN is
almost isotropy in compressibility (0.317\%) and has slightly
shear anisotropy ($A_{G}$=4.199\%). From Table III, one can see
that cubic CsCl-ReN is compression isotropy, and hexagonal ReN's
$c$ axis has much stronger incompressibility than that of $a$
axis, which is in good agreement with fractional axis compression
analysis mentioned above. The small elastic anisotropy of
NiAs-ReN, including linearized compressibility and shear
anisotropy, along with the ultra-high bulk modulus and the higher
shear modulus (or $c_{44}$), indicates that NiAs-ReN can be as a
good candidate of superhard material. Comparing $c_{44}$ with
shear modulus for hexagonal NiAs-ReN, we find that they are almost
the same, imaging that NiAs-ReN hardness can be given a consistent
estimation by the viewpoints on hardness from Teter\cite{Teter}
and Jhi \emph{et al.}\cite{Jhi}. Here, using the correlation
between shear modulus and Vickers harness given by Teter, the
estimated Vickers hardness of ReN are 33.6 GPa for CsCl-ReN and
32.1 GPa for NiAs-ReN, and 36.2 GPa for MoN-ReN. All the hardness
of ReN presented here are comparable to that of B$_{6}$O
(35$\pm$5GPa) and that of TiB$_{2}$ (33$\pm$2GPa).\cite{Teter}

The Debye temperature correlates with many physical properties of
solids, such as elastic constants, specific heat, melting
temperature and superconduction transition temperature. Since the
vibrational excitations mainly arise from acoustic vibrations at
low temperature, the Debye temperature obtained from elastic
constants is the same as that determined from specific heat
measurements. The Debye temperature can be derived by bulk
modulus, shear modulus and density $\rho$.\cite{Anderson} The
calculated Debye temperature $\Theta_{D}$s are as high as 968 K
for CsCl-ReN and 950 K for NiAs-ReN, which indicates that ReN
maybe a superconductor. So, we further estimate the
superconducting transition temperature of ReN with NiAs structure
using the modified McMillan equation,\cite{Allen} where the
electron-phonon coupling constant $\lambda$ is calculated by
density-functional perturbation theory by means of the QUANTUM
ESPRESSO package.\cite{pwscf} The calculated $\lambda$ is 1.67 and
the estimated superconducting transition temperature $T_c$ is
about 4.8 K for the screened Coulomb pseudopotential $\mu^{*}$
equal to 0.1, which agrees well with available experimental
result.\cite{Haq}
%Thus, we estimate superconductivity transition temperature $T_{c}$
%by using the BCS approximate formula
%\begin{equation}\label{BCS}\nonumber
%T_{C}=1.13\Theta_{D} \exp\left[-\frac{1}{N_{F}\lambda}\right],
%\end{equation}
%where $N_{F}$ is the density of states at the Fermi level per volume,
%and $\lambda$ is the effective electron phonon coupling parameter.

%For hexagonal crystalline, linearized compressible factor is defined as
%\begin{equation}\label{compression}\nonumber
%\frac{B_{c}}{B_{a}}=\frac{c_{33}-c_{13}}{c_{11}+c_{12}-2c_{13}}.
%\end{equation}

%\subsection{Electronic property}
To understand the correlation between the electronic properties
and the mechanical properties, we present density of states (DOS)
and the electron density distribution of ReN in the equilibrium
geometry. Electronic structure properties play an important role
in material physical properties. The total density of states (DOS)
and partial DOS of two phases of ReN are shown in Fig. 4 and Fig.
5. The obtained $N(E_{F})$ and the linear specific heat
coefficient $\gamma$ are given in Table I. The finite value
$N(E_{F})$ shows that ReN is a metal. For two phases, the
electrons from Re-\emph{d} and the N-\emph{p} states both
contribute to the DOS near the Fermi level. The peak emerging in
the lower energy region of the DOS curve mainly originates from
the localized \emph{s} states of N. The energy region just above
Fermi level is dominated by unoccupied Re \emph{d} states. The DOS
of Re-\emph{d} and N-\emph{p} are energetically degenerate from
the bottom of the valence band to the top of conduction band,
indicating the possibility of covalent bonding between Re and N
atoms. The covalent characteristic between Re and N atoms can be
confirmed by the charge density distribution. The charge density
distribution in (110) plane for CsCl-ReN and (11$\bar{2}$0) plane
for NiAs-ReN are shown in Fig. 6 (a) and (b), respectively. It is
clearly seen that a strong directional bonding exists between Re
and N in the ReN, which contributes to high bulk modulus $B$,
namely, strong incompressibility of ReN. Additionally, the
Mulliken atomic population analysis is carried out by using CASTEP
code. The obtained total charge transfer from Re to N in CsCl-ReN
(NiAs-ReN) is 0.55 (0.62), implying that the chemical bondings
between Re and N have some characteristics of ionicity. Thus our
results demonstrate that the bondings should be a mixture of
covalent, mental, and ionic attribution in ReN.

%-----------------------------------------------------Fig.------------------------

\section{Conclusion}
The structural, elastic, and electronic properties of different
structures for ReN are investigated based on first-principles
calculation under the framework of density functional theory
within local density approximation. Two competing structures,
i.e., CsCl-ReN and NiAs-ReN, are found. High valence electron
density, strong incompressibility and low elastic anisotropy
indicate that ReN is an ultra-incompressible superhard solid.
Electron structure calculation shows that ReN presents an obvious
metal feature. In particular, we obtain a superconducting
transition temperature $T_c$$\approx$4.8 K for hexagonal ReN. The
strong hybridization between metal \emph{d}-electron and nonmetal
\emph{p}-electron are observed, indicating that there is a strong
covalent bonding between Re and N. From Mulliken atomic population
analysis along with discussion of mechanical and electronic
properties, it can be concluded that the bondings in ReN should be
a mixing of metal, ionic and covalent characteristics. Our results
indicate that ReN can be used as a potential ultra-incompressible
conductor. We hope this work can stimulate the experimental
research for ReN.

 \vspace{1 mm}
%\section{Acknowledgement}
This work was supported by the National Science Foundation of
China under Grant Nos 10504036 and 90503005, the special Funds for
Major State Basic Research Project of China(973) under grant no.
2005CB623603, Knowledge Innovation Program of Chinese Academy of
Sciences, and  Director Grants of CASHIPS. Part of the
calculations were performed in the Shanghai Supercomputer Center.

\newpage

\noindent {\bf {\large {FIGURE CAPTIONS}}}

\vglue 1.0cm

\noindent {\bf {Fig.1:}} (Color online) The total energy per ReN
molecule as a function of volume (EOS).

\vglue 1.0cm

\noindent {\bf {Fig.2:}} (Color online) Volume as a function of
pressure with respect to equilibrium volume. $V_{0}$ is the
equilibrium volume.

\vglue 1.0cm

\noindent {\bf {Fig.3:}} (Color online) Fractional axis
compression as a functional of pressure. $a_{0}$, $b_{0}$, and
$c_{0}$ are lattice parameters of equilibrium volume at zero
pressure.

\vglue 1.0cm

\noindent {\bf {Fig.4:}} (Color online) Partial and total DOS for
the CsCl phase of ReN obtained from FP-LAPW calculations. Vertical
dotted lines indicate the Fermi level.

\vglue 1.0cm

\noindent {\bf {Fig.5:}} (Color online) Partial and total DOS for
the NiAs phase of ReN obtained from FP-LAPW calculations. Vertical
dotted lines indicate the Fermi level.

\vglue 1.0cm

\noindent {\bf {Fig.6:}} The valence electron density contour for
ReN obtained from FP-LAPW. Charge density is in an increment of
0.01 \emph{e}/a.u.$^{3}$ from 0.01 \emph{e}/a.u.$^{3}$ to 0.12
\emph{e}/a.u.$^{3}$. (a) (110) plane for CsCl-ReN; (b)
(11$\overline{2}$0) plane for NiAs-ReN.

\newpage
\noindent {\bf {\large {TABLE CAPTIONS}}}

\vglue 1.0cm \noindent {\bf {TABLE I:}} Equilibrium lattice
parameters,\emph{V}$_{0}$(\AA$^{3}$), \emph{a} (\AA), \emph{c}
(\AA), $c/a$, the shortest Re-N distance $d$, density $\rho$
(g/cm$^{3}$), valence electron densities $\rho_{e}$
(electrons/\AA$^{3}$), bulk modulus \emph{B}$_{0}$ and its
pressure derivative \emph{B}$^{'}_{0}$, relative total energy
\emph{E}$_{tot}$ (eV), DOS at Fermi level $N(E_{F})$
(States/eV/cell), and the linear specific heat coefficient
$\gamma$ (mJ/mol.cell.K$^{2}$). \emph{V}$_{0}$ and
\emph{E}$_{tot}$ are of per chemical f.u.

\vglue 1.0cm \noindent {\bf {TABLE II:}} Zero-pressure elastic
constants \emph{c}$_{ij}$ (GPa), the isotropic bulk modulus
\emph{B} (GPa), shear modulus \emph{G} (GPa), Young's modulus
\emph{E} (GPa), Possion's ratio $\nu$, average elastic wave
velocity $\upsilon_{m}$ (m/s), and Debye temperature $\Theta_{D}$.

\vglue 1.0cm \noindent {\bf {TABLE III:}} The bulk modulus along
the crystallographic axes \emph{a}, \emph{b}, and \emph{c}
(\emph{B}$_{a}$, \emph{B}$_{b}$, and \emph{B}$_{c}$) for ReN.
Percent elastic anisotropy for shear and bulk moduli
\emph{A}$_{G}$ (in \%), \emph{A}$_{B}$ (in \%) and compressibility
anisotropy factors \emph{A}$_{B_{a}}$ and \emph{A}$_{B_{c}}$ for
ReN obtained from LDA calculations. Here, $\emph{A}_{B_{a}}$
=$\frac{B_{a}}{B_{b}}$, $\emph{A}_{B_{c}}$ =$\frac{B_{c}}{B_{b}}$.

\newpage
%--------------------------------------Table I---------------------
%---------------
\begin{table}[htbp]
\begin{center}
\caption{\hspace{12pt} Li \textit {et al.}} \vspace{0.1cm}
%\begin{tabular}{ccccp{0.1mm}cccc} \hline \hline
\begin{tabular}{ccccccc} \hline \hline
&ZB-ReN &NaCl-ReN &CsCl-ReN &WC-ReN     &NiAs-ReN   \\
\hline

\emph{V}$_{0}$ &23.5037 &19.5258  &18.9065 &19.4775  &19.0162  \\
\emph{a}       &4.5468  &4.2743   &2.6640  &2.7476   &2.7472   \\
\emph{c}       &        &         &        &2.9786   &5.8180   \\
%%$c/a$        &        &         &        &         &2.1178   \\
\emph{d}       &1.9688  &2.1371   &2.3073  &2.1759   &2.1523   \\
$\rho$         &14.1476 &17.0299  &17.5877 &17.0721  &17.4860  \\
$\rho_{e}$     &0.5106  &0.6146   &0.6347  &0.6161   &0.6310   \\
B$_{0}$        &340   &406   &412   &433    &453    \\
B$^{'}_{0}$    &4.241 &4.729 &4.542 &4.319  &4.203  \\
E$_{tot}$      &0.444 &1.328 &1.623 &0.610  &0.0    \\
$N(E_{F})$     &-     &-     &1.220 &-      &1.674  \\
$\gamma$       &-     &-     &2.88  &-      &3.95   \\
\hline \hline
\end{tabular}
\end{center}
\end{table}

\clearpage
\newpage
%------------------------------------------------Table II-------------------------
\begin{table}[htbp]
\begin{center}
\caption{\hspace{12pt} Li \textit {et al.}} \vspace{0.1cm}
\begin{tabular}{lccccccccccc}
\hline \hline
       &       &\emph{c}$_{11}$  &\emph{c}$_{33}$  &\emph{c}$_{44}$  &\emph{c}$_{12}$  &\emph{c}$_{13}$
       &$B$    &$G$    &$E$   &$\nu$    &$\Theta_{D}$  \\
\hline

&ZB-ReN    &297 &  &-715 &356 & &336 &-255 &-1025 &- &-\\
&NaCl-ReN  &456 &  &-334 &372 & &400 &-27  &-83   &- &- \\
&CsCl-ReN  &1015 &      &164    &124   &     &421 &248  &622  &0.25  &968     \\
&WC-ReN    &640  &654   &0.55   &240   &368  &429 &53   &152  &-     &-     \\
&NiAs-ReN  &712  &897   &236    &314   &278  &450 &237  &604  &0.28  &950     \\

\hline \hline
\end{tabular}
\end{center}
\end{table}

%----------------------------------------------Table III-------------------------
\begin{table}[htbp]
\begin{center}
\caption{\hspace{12pt} Li \textit {et al.}} \vspace{0.1cm}
\begin{tabular}{lcccccccc}
\hline \hline
       &   &\emph{B}$_{a}$ &\emph{B}$_{b}$ &\emph{B}$_{c}$ &\emph{A}$_{B_{a}}$ &\emph{A}$_{B_{c}}$ &\emph{A}$_{B}$  &\emph{A}$_{G}$  \\
\hline

&CsCl-ReN            &1264     &  &       &1   &        &0      &11.55  \\

%&$\delta_{1}$-ReN    &1065.4      &2112.3 &1   &1.9827  &0.733  &97.42  \\

&NiAs-ReN           &1237      & &1627 &1   &1.3155   &0.317  &4.199 \\

%%&MoN-ReN    &1242    &  &1606 &1   &1.2934  &0.294  &0.171  \\

\hline \hline
\end{tabular}
\end{center}
\end{table}
%-----------------------------------------------------------------
\clearpage
\newpage
%---------------------------------------Fig.1----------------------
\begin{figure}[htbp]
 \includegraphics[width=7.0 cm, angle=0,clip]{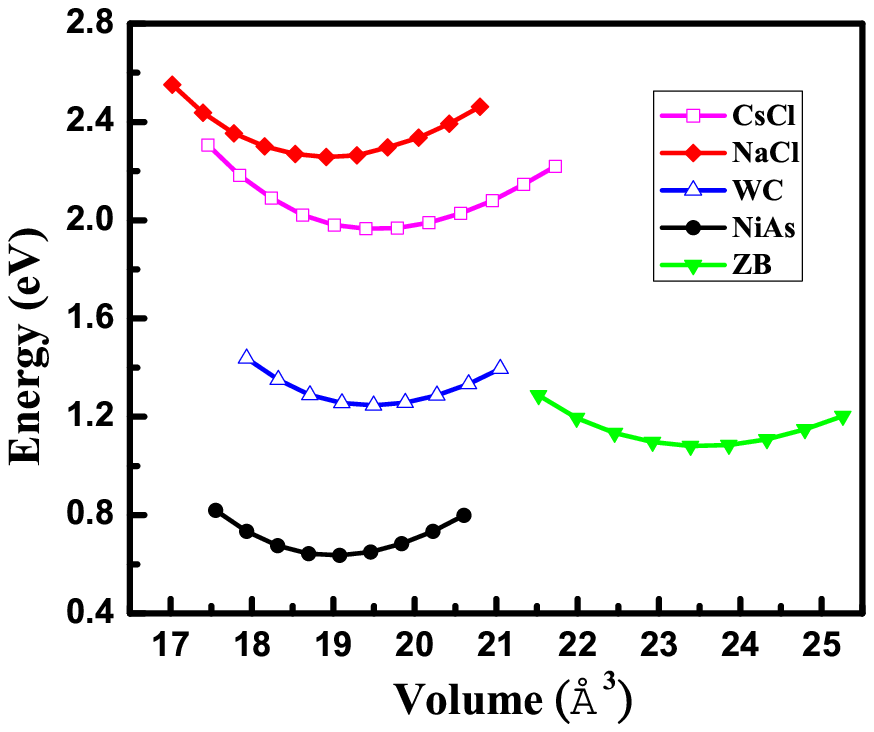}
 \caption{\hspace{12pt} Li \textit {et al.}}
\end{figure}

\clearpage
\newpage
%---------------------------------------Fig.2---------------

\begin{figure}[htbp]
 \includegraphics[width=8.0 cm, angle=0,clip]{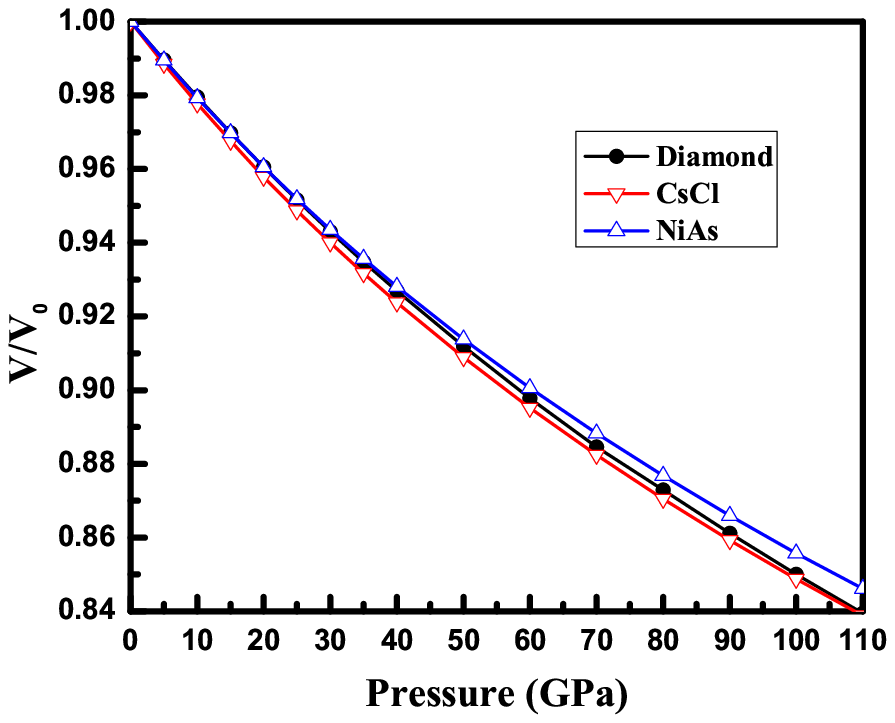}
 \caption{\hspace{12pt} Li \textit {et al.}}
\end{figure}

\clearpage
\newpage
%---------------------------------------Fig.3----------------------
\begin{figure}[htbp]
 \includegraphics[width=8.0 cm, angle=0,clip]{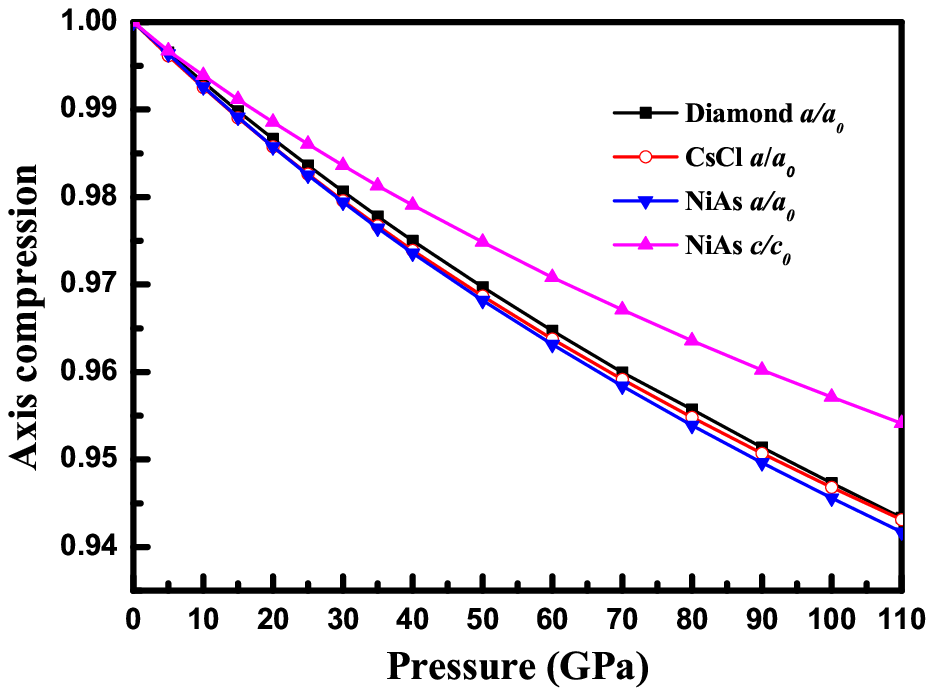}
 \caption{\hspace{12pt} Li \textit {et al.}}
\end{figure}

\clearpage
\newpage
%---------------------------------------Fig.4----------------------
\begin{figure}[htbp]
 \includegraphics[width=7.0 cm, angle=0,clip]{fig4.eps}
 \caption{\hspace{12pt} Li \textit {et al.}}
\end{figure}

\clearpage
\newpage
%---------------------------------------Fig.5-----------------------
\begin{figure}[htbp]
 \includegraphics[width=7.0 cm, angle=0,clip]{fig5.eps}
 \caption{\hspace{12pt} Li \textit {et al.}}
\end{figure}

\clearpage
\newpage
%---------------------------------------Electron density---------------------
\begin{figure}[htbp]
 \includegraphics[width=7.0 cm, angle=0,clip]{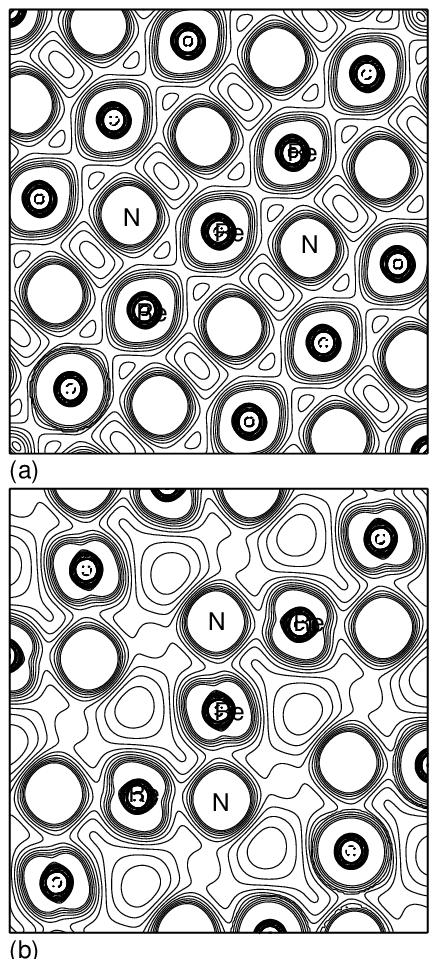}
 \caption{\hspace{12pt} Li \textit {et al.}}
\end{figure}


\begin{thebibliography}{000}
\bibitem{Horvath-Bordon}
E. Horvath-Bordon, R. Riedel, A. Zerr, P. F. McMillan, G.
Aufermann, Y. Prots, W. Bronger, R. Kniep, and P. Kroll, Chem.
Soc. Rev. {\bf 35}, 987 (2006).

\bibitem{PMcMillan}
P. F. McMillan, Nat. Mater. 1, 19 (2002)

\bibitem{Zerr}
A. Zerr, G. Miche, and R. Riedel, Nat. Mater. 2, 185 (2003).

\bibitem{Crowhurst}
J. C. Crowhurst, A. F. Goncharov, B. Sadigh, C. L. Evans, P. G.
Morrall, J. L. Ferreira, and A. G. Nelson, Science {\bf 311}, 1275
(2006).

\bibitem{Young}
 A. F. Young, C. Sanloup, E. Gregoryanz, S. Scandolo, R. J. Hemley, and H. K. Mao, Phys. Rev. Lett. {\bf 96}, 155501 (2006).

\bibitem{Soignard}
E. Soignard, O. Shebanova, and P. F. McMillan, Phys. Rev. B {\bf
75}, 014104 (2007).

\bibitem{Bull}
C. L. Bull, P. F. McMillan, E. Soignard, and K. Leinenweber, J.
Solid State Chem. {\bf 177}, 1488 (2004).

\bibitem{Kaner}
R. B. Kaner, J. J. Gilman and S. H. Tolbert. Science {\bf 308},
1268 (2005).

\bibitem{Mattesini}
M. Mattesini, R. Ahuja, and B. Johansson, Phys. Rev. B {\bf 68},
184108 (2003).

\bibitem{Young-1}
A. F. Young, J. A. Montoya, C. Sanloup, M. Lazzeri, E. Gregoryanz,
and S. Scandolo, Phys. Rev. B {\bf 73}, 153102 (2006).

\bibitem{Chen}
 Z. W. Chen, X. J. Guo, Z. Y. Liu, M. Z. Ma, Q. Jing, G. Li, X. Y. Zhang, L. X. Li,
  Q. Wang,Y. J. Tian, and R. P. Liu, Phys. Rev. B {\bf 75}, 054103 (2007).

\bibitem{Yu}
 R. Yu, Q. Zhan, and L-C. De Jonghe, Angew. Chem. Int. Ed. {\bf 46}, 1136 (2007).

\bibitem{Wu}
 Z. Wu, X. Hao, X. Liu, and J. Meng, Phys. Rev. B {\bf 75}, 054115 (2007).

\bibitem{Montoya}
J. A. Montoya, C. Sanloup, E. Gregoryanz, and S. Scandolo, Appl.
Phys. Lett. {\bf 90}, 011909 (2007).

\bibitem{Fan}
C. Z. Fan, S. Y. Zeng, L. X. Li, Z. J. Zhan, R. P. Liu, W. K.
Wang, P. Zhang, and Y. G. Yao, Phys. Rev. B {\bf 74}, 125118
(2006).

\bibitem{Kanoun}
M. B. Kanoun, S. Goumri-said, and M. Jaouen, Phys. Rev. B {\bf
76}, 134109 (2007).

\bibitem{Haq}
A. ul. Haq and O. Meyer, J. Low Temp. Phys. {\bf 50}, 123 (1983).

\bibitem{Isaev}
E. I. Isaev, S. I. Simak, I. A. Abrikosov, R. Ahuja, Yu. Kh.
Vekilov, M. I. Katsnelson, A. I. Lechtenstein, and B. Johansson,
J. Appl. Phys. {\bf 101}, 123519 (2007).

\bibitem{MoN}
The $\delta$-MoN (space group $P6_{3}/mc$) type is also considered
in our research. It is found that $\delta$-MoN type is also stable
mechanically and has a strong incompressibility (bulk modulus $B$,
451 GPa; shear modulus $G$, 267 GPa). Symmetry analysis show that
the MoN-ReN structure is a slight distortion of the supercell
(2$\times$2$\times$1) of NiAs-ReN structure. The unit cell
(\emph{a}=5.4981 and \emph{c}=5.8095 \AA) of MoN-ReN contains
eight f.u. with eight Re atoms occupying $2a$ (0, 0, 0.5481) and
$6c$ (0.4996, 0.5004, 0.5472) sites and eight N atoms holding $2b$
($\frac{1}{3}$, $\frac{2}{3}$, 0.7994) and $6c$ (0.1667, 0.8333,
0.2986) sites. The NiAs-ReN and MoN-ReN have the very similar
structural parameters, which result in the almost same total
energy per f.u. (differing only by 5 meV) and the similar
mechanical and electronic properties. So, the results of
$\delta$-MoN type is not given in the text.

%\bibitem{Chung}
%H-y.Chung, M.B.Weinberger, J.B.Levine, A.Kavner, J-M.Yang, S.H.Tolbert, and R.B.Kaner, Science {\bf 316}, 436(2007).
\bibitem{Allen}
P. B. Allen and R. C. Dynes, Phys. Rev. {\bf12}, 905 (1975).

\bibitem{APW}
 E. Sj\"{o}stedt, L. Nordstr\"{o}m, and D. J. Singh, Solid State Commun. {\bf 114}, 15 (2000);
 G. K. H. Madsen, P. Blaha, K. Schwarz, E. Sj$\ddot{o}$stedt, and L. Nordstr$\ddot{o}$m, Phys. Rev. B {\bf 64}, 195134 (2001).

\bibitem{wien2k}
 P. Blaha, K. Schwarz, G. Madsen, D. Kvasnicka, and J. Luitz, computer code Wien2k, an augmented plane wave plus local orbitals
program for calculating crystal properties, Karlheinz
Schwarz,Technische Universit$\ddot{s}$t Wien, Austria, 2001.

\bibitem{Segall}
M. D. Segall, P. J. D. Lindan, M. J. Probert, C. J. Pickard, P. J.
Hasnip, S. J. Clark, M. C. Payne, J. Phys.: Cond. Matt. {\bf 14},
2717 (2002).

\bibitem{BFGS}
B. G. Pfrommer, M. Cote, S. G. Louie, and M. L. Cohen, J. Comput.
Phys. 131, 133 (1997).

\bibitem{Hill}
R. Hill, Proc. Phys. Soc. London {\bf 65}, 349 (1952).

\bibitem{eos}
F. Birch, Phys.Rev. {\bf 71}, 809 (1947); J. -P. Poirier,
Introduction to the physics of the Earth's Interior, Cambridge
University Press, Cambridge, 2000.

\bibitem{Neumann}
G. S-Neumann, L. Stixrude, and R. E. Cohen, Phys. Rev. B {\bf 60},
791 (1999).

\bibitem{Chung}
H. Chung, A. B. Weinberger, J. B. Levine, A. Kavner, J. Yang, S.
H. Tolbert, and R. B. Kaner, Science {\bf 316}, 436 (2007).

\bibitem{Gilman}
J. J. Gilman, R. W. Cumberland, and R. B. Kaner, Int. J. Refrac.
Mat \& Hard Mat.  {\bf 24}, 1-5 (2006).

\bibitem{diamond}
F. Occelli, D. L. Farber, and R. L. Toullec, Nat. Mater. {\bf 2},
151 (2003); J. C. Zheng, Phys. Rev. B {\bf 72}, 052105 (2005).

\bibitem{optimize}
All calculated parameters using PW-PP method agree well with those
given by using FP-LAPW method within 1.0\%, showing that the PW-PP
method can be used to calculate the elastic properties of ReN.

\bibitem{Born}
M. Born and K. Huang, Dynamical Theory of Crystal Lattices
(Clarebdom, Oxford, 1956).

\bibitem{Ravindran}
P. Ravindran, L. Fast, P. A. Korzhavyi, B. Johansson, J. Wills,
and O. Eriksson, J. Appl. Phys. {\bf 84}, 4891 (1998).

\bibitem{Teter}
D. M. Teter, MRS. Bull. {\bf 23}, 22 (1998).

\bibitem{Jhi}
S-H. Jhi, J. Ihm, S. G. Louie, and M. L. Cohen, Nature
{\bf399},132 (1999).

\bibitem{Anderson}
O. L. Anderson, J. Phys. Chem. Solids {\bf 24}, 909 (1963).
% Debye temperature

%W. L. McMillan, Phys. Rev. 167, 331 (1968).
\bibitem{pwscf}
P. Giannozzi et al., http://www.quantum-espresso.org.

\end{thebibliography}
\end{document}